\documentclass[a4paper,12pt]{article}
\usepackage[utf8]{inputenc}
\usepackage{cancel}
\usepackage{ulem}
\usepackage{amsfonts}
\usepackage{amssymb}
\usepackage{graphicx}
\usepackage{amsmath}
\usepackage{enumerate}
\usepackage{mathtools}
\usepackage{subfig}
\usepackage{color}
\usepackage{tikz}\usetikzlibrary{calc}
\usepackage{float}
\usepackage{here}
\usepackage{cite}
\usepackage{mathrsfs}
\usepackage{float,epsfig}
\usepackage{dcolumn}

\usepackage{graphicx}
\usepackage{bm}
\usepackage{amsmath,amssymb,amsthm}
\usepackage[colorlinks=true,linkcolor=blue,citecolor=red]{hyperref}
\newcommand{\be}{\begin{equation}}
\newcommand{\ee}{\end{equation}}
\newcommand{\bea}{\setlength\arraycolsep{2pt} \begin{eqnarray}}
\newcommand{\eea}{\end{eqnarray}}

\def\0{{\sst{(0)}}}
\def\1{{\sst{(1)}}}
\def\2{{\sst{(2)}}}
\def\3{{\sst{(3)}}}
\def\4{{\sst{(4)}}}
\def\5{{\sst{(5)}}}
\def\6{{\sst{(6)}}}
\def\7{{\sst{(7)}}}
\def\8{{\sst{(8)}}}
\def\sst#1{{\scriptscriptstyle #1}}

\makeatletter \@addtoreset{equation}{section}

\usepackage{multirow}
\begin{document}
%

\title{\textbf{\Large On M-theory on Real Toric Fibrations }}
\author{ A. Belhaj$^1$, H. Belmahi$^1$, M. Benali$^1$, S-E. Ennadifi$^2$, Y.
Hassouni$^1$, Y. Sekhmani$^1$\footnote{Authors in alphabetical order.} \hspace*{-8pt} \\
{\small $^1$ Equipe des Sciences de la mati\`ere et du rayonnement, ESMaR,
D\'epartement de Physique}\\
{\small Facult\'e des Sciences, Universit\'e Mohammed V de Rabat, Morocco}\\
{\small $^2$ LHEP-MS, D\'epartement de Physique, Facult\'e des Sciences}\\
{\small Universit\'e Mohammed V de Rabat, Morocco} }
\maketitle

\begin{abstract}
{\noindent}

Borrowing ideas from elliptic complex geometry, we approach M-theory
compactifications on real fibrations. Precisely, we explore real toric
equations rather than complex ones exploited in F-theory and related dual
models. These geometries have been built by moving real toric manifolds over
real bases. Using topological changing behaviors, we unveil certain data
associated with gauge sectors relying on affine Lie symmetries. \newline
\end{abstract}

%
%

\newpage 

\newpage

\section{Introduction}

Complex algebraic geometry has been considered as a crucial tool not only in
mathematics but also in physics \cite{111}. In particular, it has been used
in modern physics and related topics. It concerns building of compact complex
manifolds needed for string compactification mechanisms \cite{1}. A special
emphasis has been put on Calabi-Yau geometries supported by the industry of
stringy models providing certain semi-realistic theories going beyond
standard models. Among others, the elliptic fibrations of Calabi-Yau
manifolds have been extensively investigated in connection with F-theory and
stringy dual models. It is recalled that F-theory, which is a 12-dimensional
theory, was proposed after the discovery of dualities between superstring
theories in $10$-dimensions \cite{1}. This theory has been considered as an
alternative way to approach the strong coupling limits of certain
superstring theories from higher dimensions being not discovered in M-theory
scenarios\cite{2}. Exploiting the $SL(2,\mathbb{Z})$ symmetry, the
compactification of such a theory on the torus $\mathbb{T}^{2}$ has been
studied. In a specific limit of the $\mathbb{T}^{2}$ moduli space, this can
produce type IIB superstring theory in $10$-dimensions \cite{1}. This has
been established by identifying the $\mathbb{T}^{2}$ shape complex moduli
with a complex scalar field formed by combining the axion $\chi $ and the
dilaton $\phi $ fields. Lower dimensional F-theory models have been built
using the elliptic fibrations on complex base manifolds, providing many
supersymmetric vacuum solutions \cite{3,4}. In particular, the associated
gauge theories have been geometrically engineered from complex elliptic
fibrations of Calabi-Yau manifold spaces \cite{5,6,7,12,13}. However, the
torus involves another parameter controlling its size which has been frozen
in F-theory building models.  This size  has been frozen because it is not invariant  under $SL(2,\mathbb{Z})$ symmetry
describing the self duality   of type  IIB superstring.  In such activities, it has been considered as a non dynamical quantity.   At this level, one could naturally ask certain
questions about such a real parameter using elliptic real fibrations in
stringy compactification scenarios. \newline
The aim of this work is to mimetic the complex elliptic fibrations by
introducing toric fibrations relying on  a real algebraic geometry. A close
inspection shows that M-theory could be considered as a possible candidate
where real toric fibrations could find a place. Concretely, we first explore
equations describing real fibration manifolds, rather than  complex ones
exploited either in F-theory or in dual theories. Then, we study the
M-theory compactification on such real fibration geometries and propose a
conjecture in terms of heterotic superstring models. These geometries have
been constructed by moving real toric manifolds over real bases. Using
topological changing behaviors, we unveil certain data associated with gauge
sectors relying on affine Lie symmetries.\newline
The organization of this work is as follows. In section 2, we deal with
M-theory on $\mathbb{T}^{2}$ using a real algebraic geometry. In section 3,
we explore real toric fibrations to approach a M-theory compactification
down to eight dimensions and make contact with Higgising matter contents.
Section 4 concerns an extended compactification of such real elliptic
fibrations. The last section is devoted to concluding remarks and
conclusions.

\section{ M-theory on real torus}

In this section, we investigate M-theory on a real algebraic geometry of $%
\mathbb{T}^{2}$. At low energies, this theory is described by eleven
dimensional supergravity involving certain bosonic fields coupled to
M-branes. Before going ahead, it is recalled that the moduli space of $%
\mathbb{T}^{2}$ involves two different kinds of parameters which control the
shape and the size associated with the metric deformations. A nice way to
understand such a moduli space is to exploit string theory
compactifications. Indeed, one uses the following parameter space
\begin{equation}
\mathcal{M}(\mathbb{T}^{2})=\frac{SL(2)}{U(1)}\times \frac{SL(2)}{U(1)}
\end{equation}%
where the first factor $\frac{SL(2)}{U(1)}$ can be viewed as the complex
structure and the other one corresponds to the K\"{a}hler structure and the
stringy B-field. Removing such a B-field, the reduced factor corresponds to
the size deformation, being frozen in certain elliptic Calabi-Yau
compcatifiactions relying on the complex geometry \cite{3,4,5,6,7}. Here,
however, we would like to elaborate real toric fibration compactifications
in M-theory by combining the real region of the $\mathbb{T}^{2}$ moduli
space and real toric geometry bundles.
Indeed, an examination shows that $\mathbb{T}^{2}$ involves a real algebraic
representation given by the following equation
\begin{equation}
\left( R-\sqrt{x_{2}^{2}+x_{3}^{2}}\right) ^{2}+x_{1}^{2}=r^{2},
\end{equation}%
where $x_{1}$, $x_{2}$ and $x_{3}$ are real coordinates of the 3-dimensional
non compact flat space \cite{14}. $R$ and $r$ are real parameters
corresponding to the involved fibred circles having  different lengths with the
condition $R>r$ needed to provide a torus manifold. For $R=2$ and $r=1$,
this real geometry is illustrated in figure 1.
\begin{figure}[h]
\begin{center}
\includegraphics[scale=0.2]{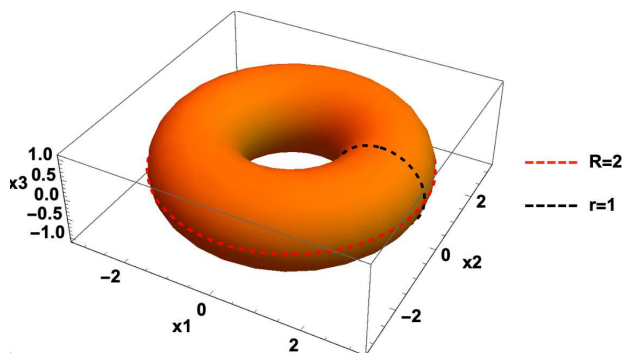}
\end{center}
\caption{Graphic representation of two dimensional torus for $R=2$ and $r=1$%
. }
\label{tsch0}
\end{figure}
It is noted that $R$ and $r$ are now real parameters controlling the area of
$\mathbb{T}^{2}$.
This geometry can be parameterized by the following real relations
\begin{eqnarray}
x_{1} &=&r\sin \theta _{1},  \notag \\
x_{2} &=&(R+r\cos \theta _{1})\sin \theta _{2},  \label{223} \\
x_{3} &=&(R+r\cos \theta _{1})\cos \theta _{2},  \notag
\end{eqnarray}%
where $\theta _{1}$ and $\theta _{2}$ are the toric coordinates of the torus
constrained by $0\leq \theta _{i}\leq 2\pi $ $(i=1,2)$. In this geometry,
the metric  line element can be written as
\begin{equation}
ds^{2}=g_{ab}d\theta ^{a}d\theta ^{b},\qquad a,b=1,2.
\end{equation}%
This metric $g_{ab}$ is given by
\begin{equation}
g_{ab}=\sum_{i,j=1}^{3}\delta _{ij}\frac{\partial x^{i}}{\partial
\theta ^{a}}\frac{\partial x^{j}}{\partial \theta ^{b}}=%
\begin{pmatrix}
r^{2} & 0 \\
0 & (R+r\cos \theta _{1})^{2}%
\end{pmatrix}%
,\qquad i,j=1,2,3,
\end{equation}%
and where $\delta _{ij}$ is the Kronecker symbol. The area of $\mathbb{T}^{2}
$ can be computed using the following integration equation
\begin{equation}
\mathit{A}(\mathbb{T}^{2})=\iint_{\mathbb{T}^{2}}\sqrt{|g|}d\theta
_{1}d\theta _{2},  \label{aire}
\end{equation}%
where $g$ is the determinant of the induced metric. Using the equation (\ref%
{aire}), one obtains
\begin{equation}
\mathit{A}(\mathbb{T}^{2})=\int_{0}^{2\pi }d\theta _{1}\int_{0}^{2\pi
}r(R+r\cos \theta _{1})d\theta _{2}=4\pi ^{2}rR.
\end{equation}%
It follows that this real analysis of $\mathbb{T}^{2}$ involves only size
parameter contributions controlling the area. It is clear that this road
does not provide any complex scalar field in M-theory compactifications.
However, it could give only a real scalar field which can be identified with
the size of $\mathbb{T}^{2}$. A close inspection shows that one can consider
the dilaton scalar filed as a possible candidate. Forgetting about the full
desired spectrum of the compactified M-theory, it has been observed that $R$
and $r$ should be considered as relevant parameters to approach such a field
content in lower dimensions. We expect that these lower compactifications
could bring extra information on such real toric fibrations in the
compactification of M-theory. By mimicking the analysis explored in the
complex algebraic geometry of F-theory stringy models, we attempt to
investigate the involved compactifications by varying the size parameters of
$\mathbb{T}^{2}$ over real base manifolds, rather than complex bases
exploited in known higher dimensional theories, including superstringy
models and F-theory.

\section{Eight dimensional models}

Borrowing ideas from known complex compactifications, the lower dimensional
models from real toric fibrations in compactifications of M-theory can be
obtained by varying circle length parameters over $d$ -dimensional real
bases by freezing the shape parameter of $T^{2}$ corresponding to the first
factor of $M(T^{2})$. This could give models in $(9-d)$ dimensions.\newline
To see how this works in practice, we deal with the case $d=1$ associated
with the circle and the interval compactifications. However, to get a
non-trivial geometry and motivated by stringy known results including Ho\v{r}%
ava-Witten models, we deal with only an interval compactification. A priori,
there are many ways to construct the resulting desired geometry. A way is to
take the following real algebraic fibration equation
\begin{equation}
\left( R(\eta )-\sqrt{x_{2}^{2}+x_{3}^{2}}\right) ^{2}+x_{1}^{2}=r(\eta
)^{2},  \label{31}
\end{equation}%
where $\eta $ indicates the coordinate of the interval real base  being constrained
by $0\leq \eta \leq \pi $. It is recalled that this real base space geometry
could be built from a circle $S^{1}$ defined by
\begin{equation}
\eta =\eta +1,
\end{equation}%
up to a $Z_{2}$ orbifold symmetry acting as follows
\begin{equation}
\eta \rightarrow -\eta .
\end{equation}%
Roughly speaking, non-trivial real fibrations can be handled by considering
generic length functions. However, we expect that periodicity, linearity and
quadratic functions are useful mathematical features providing simple tools
to visualize such real toric fibrations. To make contact with known models,
we propose the following specific variations
\begin{eqnarray}
R(\eta ) &=&R\sin \eta , \\
r(\eta ) &=&r.
\end{eqnarray}%
In this way, the equation (\ref{223}) becomes
\begin{eqnarray}
x_{1}(\eta ) &=&r\sin \theta _{1},  \notag \\
x_{2}(\eta ) &=&(R\sin \eta +r\cos \theta _{1})\sin \theta _{2},  \label{n}
\\
x_{3}(\eta ) &=&(R\sin \eta +r\cos \theta _{1})\cos \theta _{2}.  \notag
\end{eqnarray}%
The associated geometry for specific points on the interval is illustrated
in figure 2.
\begin{figure}[h]
\begin{center}
\includegraphics[scale=0.2]{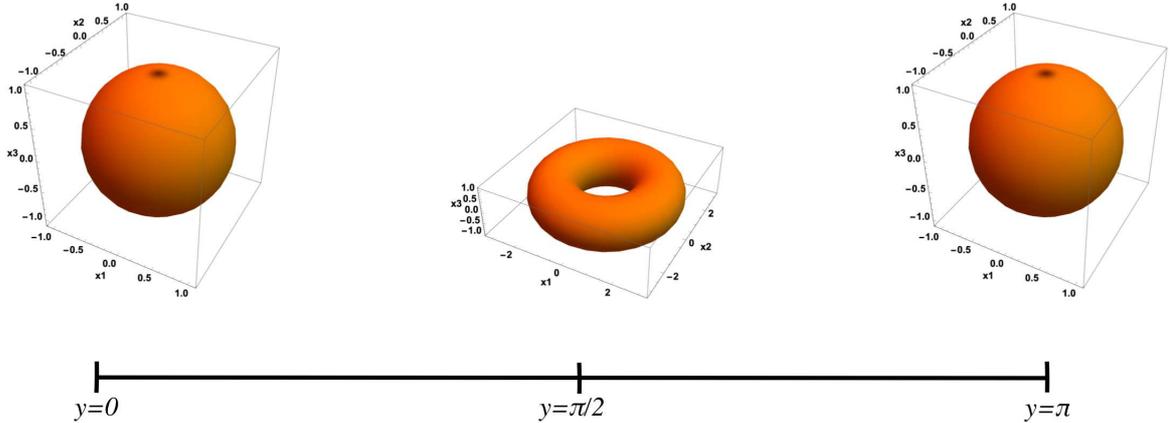}
\end{center}
\caption{Fibration behaviors on the interval $[0,\protect\pi ]$.}
\label{tsch0}
\end{figure}
It is remarked that a real $2$-sphere appears at the end of the interval $%
\eta =0$ and $\eta =\pi $ parameterized by
\begin{eqnarray}
x_{1}(\eta ) &=&r\sin \theta _{1},  \notag \\
x_{2}(\eta ) &=&r\cos \theta _{1}\sin \theta _{2}, \\
x_{3}(\eta ) &=&r\cos \theta _{1}\cos \theta _{2},  \notag \\
R(\eta ) &=&0.  \notag
\end{eqnarray}%
Its area can be calculated from the following tensor metric
\begin{equation}
g_{ab}=%
\begin{pmatrix}
r^{2} & 0 \\
0 & r^{2}\cos ^{2}\theta _{1}%
\end{pmatrix}%
.
\end{equation}%
Indeed, it is given by
\begin{equation}
\mathit{A}(\mathbb{S}^{2})=\iint_{\mathbb{S}^{2}}\sqrt{|g|}d\theta
_{1}d\theta _{2}=4\pi r^{2}.
\end{equation}%
It follows from figure 2 that $\mathbb{T}^{2}$ changes its geometry
producing non-trivial fibers over the real interval. We will see that the
area of such a real fibration is a function of the base coordinate $\eta $.
Similar ideas, in connections with material physics, have been discussed in
\cite{15}. Concretely, the area can be obtained via the following relation
\begin{equation}
\mathit{A}(\eta )=\int_{0}^{2\pi }\int_{-\pi +u(\eta )}^{\pi -u(\eta )}\sqrt{%
|g(\eta )|}d\theta _{1}d\theta _{2},
\end{equation}%
where $g(\eta )$ is now the determinant of the fibration metric
\begin{equation}
g_{ab}(\eta )=%
\begin{pmatrix}
r^{2} & 0 \\
0 & (R\sin \eta +r\cos \theta _{1})^{2}%
\end{pmatrix}%
.
\end{equation}%
It is noted that the function $u(\eta )$ has been introduced as a generic
function carrying information of the $\theta _{1}$ dependence of the
fibration. This idea has been inspired by the torus parametrization showing
that $R$ and $r\cos \theta _{1}$ should be treated on an equal footing.
After integral calculi, we obtain the following varying function area
\begin{equation}
A(\eta )=2\pi \left[ Rr\sin \eta \left( 2\pi -2u(\eta )\right) +2r^{2}\sin
u(\eta )\right] .
\end{equation}%
A simple observation shows that this area splits into two different parts
\begin{equation}
A(\eta )=a_{1}(\eta )A(\mathbb{T}^{2})+a_{2}(\eta )A(\mathbb{S}^{2}),
\end{equation}%
where $A(\mathbb{T}^{2})$ and $A(\mathbb{S}^{2})$ are the area of $\mathbb{T}%
^{2}$ and $\mathbb{S}^{2}$, respectively. However, $a_{1}(\eta )$ and $%
a_{2}(\eta )$ are real fibration functions given by
\begin{eqnarray}
a_{1}(\eta ) &=&\frac{\sin \eta }{\pi }\left( \pi -u(\eta )\right) , \\
a_{2}(\eta ) &=&\sin u(\eta ).
\end{eqnarray}%
An examination reveals that the variation of the fiber appearing in figure 2
can be assured by the following fibration function
\begin{equation}
u(\eta )=\arccos \left( \sin \eta \right) .
\end{equation}%
From the figure 2, we observe that the fibration generates a smoothness
breakdown corresponding to the appearance of a real 2-sphere at the ends of
the interval real base given by $\eta =0$ and $\eta =\pi $. This produces
two different relevant topological behaviors. The first one is singular
associated with the limits
\begin{eqnarray}
R(\eta ) &\rightarrow &0,  \notag \\
a_{1}(\eta ) &\rightarrow &0, \\
A(\eta ) &\rightarrow &A(\mathbb{S}^{2}),  \notag
\end{eqnarray}%
occurring at the ends of the interval $\eta =0,\pi $. While, the second one
is a smooth torus behavior corresponding to
\begin{eqnarray}
R(\eta ) &\rightarrow &R,  \notag \\
a_{2}(\eta ) &\rightarrow &0, \\
A(\eta ) &\rightarrow &A(\mathbb{T}^{2}),  \notag
\end{eqnarray}%
being localized at the middle of the interval $\eta =\frac{\pi }{2}$. These
behaviors relay on the Euler characteristic $\chi (\mathbb{T}^{2})=0$ and $%
\chi (\mathbb{S}^{2})=2$, respectively.
It is recalled that the Euler characteristic appears in many places in
string theory and related topics including Calabi-Yau compactifications.
Among others, it takes places in intersecting geometries used in the brane
realization of gauge theories from complex algebraic compactifications \cite{16}. For certain Calabi-Yau manifolds, these intersecting geometries are
linked to Lie symmetries \cite{17}. Based on this remark, the above singular
behaviors could be worked out to supply possible connections with certain
known theories up to some assumptions.

Similarities with such theories push one to unveil hidden sectors. Indeed,
singular behaviors of such real fibrations could be linked to Lie
symmetries. This road deserves more ideas of reflections. It is worth
noting that many interesting results in string theory and related models
have been discovered from the metric and the topological properties of
compact manifolds using the geometric engineering method \cite{200}.
Motivated by such activities, we could bring certain information on the
gauge sector. This invisible sector could be approached by combining certain
topological and geometrical pictures associated with singular behaviors, the
Euler characteristic changing, and the torus behaviors in terms of gluing
spheres from a weighted sum. To reach the goal that we are looking for, we
engineer the fiber as a 2-dimensional real space $\mathbb{M}^{2}$ by gluing $%
(n+1)$ real 2-spheres. The adjacent 2-spheres intersect on a disk. However,
we consider the limit where such a disk shrinks to a point. This geometrical
configuration has been extensively studied in connections with material
physics. Motivated by weighted sizes explored in such a physics \cite{21},
we represent the manifold $\mathbb{M}^{2}$ by a weighted connected sum of $%
(n+1)$ real 2-spheres given by
\begin{equation}
\mathbb{M}^{2}=\sum_{p=0}^{n}\omega _{p}\mathbb{S}_{p}^{2},  \label{inter}
\end{equation}%
where $\omega _{p}$ is the weight of $p$-th 2-sphere. A simple configuration
is illustrated in figure 3, by gluing only two 2-spheres.

\begin{figure}[h]
\begin{center}
\includegraphics[scale=0.2]{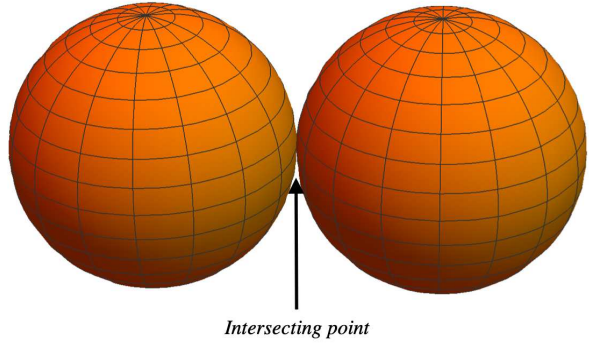}
\end{center}
\caption{Gluing spheres.}
\label{tsch0}
\end{figure}
 According to many  works including  \cite{S1,S2},  it is  noted   that the self intersection of $\mathbb{M}^2$ is linked to the
Euler characteristic via the relation
\begin{eqnarray}
\chi(\mathbb{M}^2)=-\mathbb{M}^2\cdot \mathbb{M}^2.
\end{eqnarray}
Using the equation(\ref{inter}), this can be re-written as
\begin{eqnarray}
\chi(\mathbb{M}^2)= -\sum_{q=0}^n \sum_{p=0}^n \omega_q\omega_p I_{pq}
\end{eqnarray}
where $I_{pq}$ denotes the intersection number between $\mathbb{S}_p^2$ and $%
\mathbb{S}_q^2$
\begin{eqnarray}
I_{pq}= \mathbb{S}_p^2.\mathbb{S}_q^2.
\end{eqnarray}
The vanishing of the Euler characteristic, $\chi(\mathbb{M}^2)=0$,
corresponding to the torus behavior, is assured by the following condition
\begin{equation}
\sum_{q=0}^n \omega_q\left(\sum_{p=0}^n\omega_p I_{pq}\right)=0.
\end{equation}
For generic $\omega_q$ weights, this condition can be solved by
\begin{eqnarray}
\sum_{p=0}^n\omega_pI_{pq}=0.
\end{eqnarray}
 It is noted that this constraint shears similarities with  the toric  realisation of  local  Calabi-Yau  three-folds with elliptic singularities. In particular, this can represent the  compact part   used to remove the associated singularities where the entries of  the matrix  $I_{pq}$  provide  the intersection  numbers of the involved blowing  up cycles \cite{7,12}
A priori, there could be many other ways to verify such a condition.
However, a possible one is to exploit results explored in the theory of Lie
symmetries \cite{22}. Indeed, we propose the following solution
\begin{eqnarray}
I_{pq}&=&- K_{pq} , \\
\omega_p&=&\delta_p,
\end{eqnarray}
where $K_{pq}$ and $\delta_p$ are the Cartan matrices and the Dynkin weights
of affine Lie symmetries, respectively. In this context, $n$ is identified
with the involved rank. An examination shows that the matrix $I_{pq}$ should
be symmetric which requires a particular choice for such symmetries. They
are identified with affine simply laced ones. The graphic representation of
these symmetries are given in figure 4.

\begin{figure}[h]
\begin{center}
\includegraphics[scale=0.6]{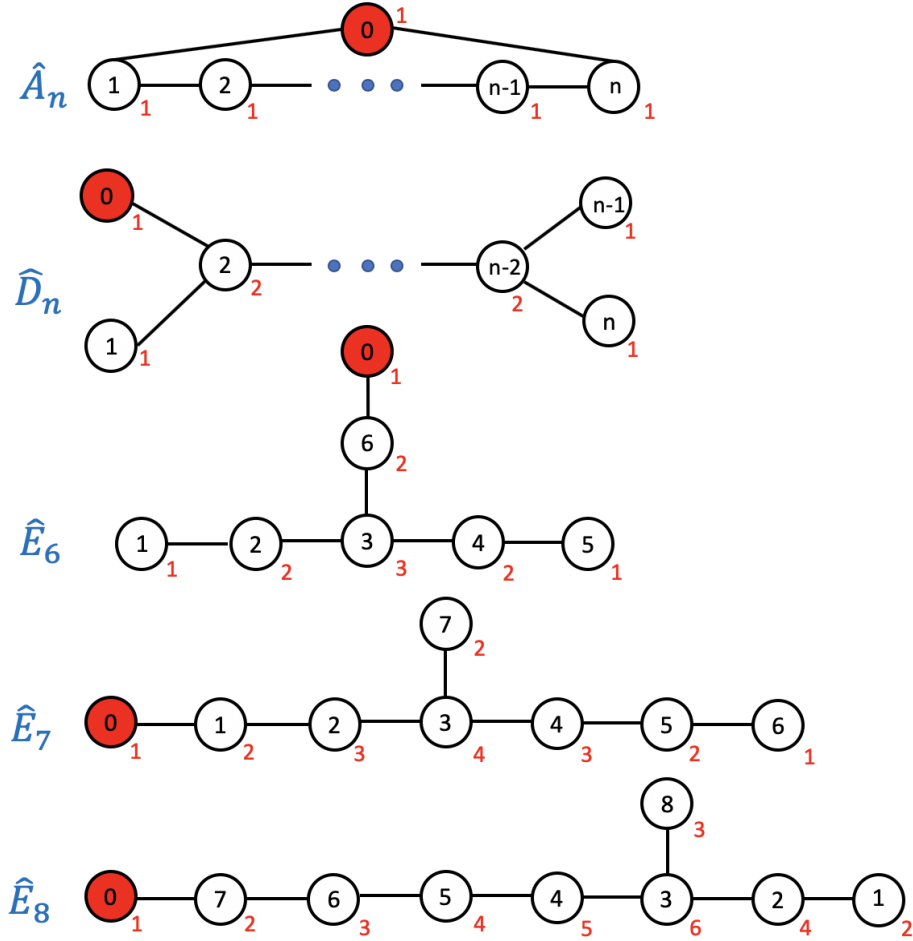}
\end{center}
\caption{ADE affine Lie symmetries with Dynkin weights given in red color.
The red vertex represents the affine one.}
\label{tsch0}
\end{figure}

In the present context, the above sum can be reduced to
\begin{equation}
\mathbb{M}^{2}=\mathbb{S}_{0}^{2}+\sum_{p=1}^{n}\delta _{p}\mathbb{S}%
_{p}^{2},
\end{equation}%
where we have taken $\delta _{0}=1$ which corresponds to the affine root.
However, the remaining values of $\delta _{p}$ ($p\neq 0)$ fix the desired
affine Lie symmetry. In this language, the torus behavior is mapped to the
imaginary root of such a Lie symmetry given by
\begin{equation}
\beta =\alpha _{0}+\sum_{p=1}^{n}\delta _{p}\alpha _{p}
\end{equation}%
where $\alpha _{p}$ are the involved simple roots. In this geometric
picture, the first 2-sphere $\mathbb{S}_{0}^{2}$ corresponds to the affine
simple root which will carry singular behavior information. Indeed, the area
of $\mathbb{M}^{2}$ takes the following form
\begin{equation}
A(\mathbb{M}^{2})=4\pi (r_{0}^{2}+\sum_{p=1}^{n}\delta _{p}r_{p}^{2})
\end{equation}%
where now $\delta _{p}$ can be regarded as radius scaling factors. It is
remarked that the torus behavior appearing at the middle of the interval
corresponds to the following size constraints
\begin{eqnarray}
r_{0} &\neq &0, \\
r_{p} &\neq &0\qquad p=1,\ldots ,n.  \notag
\end{eqnarray}%
However, these real parameters should satisfy the following relation
\begin{equation}
r_{0}^{2}+\sum_{p=1}^{n}\delta _{p}r_{p}^{2}=Rr.
\end{equation}%
In this way, the singular behaviors located at the interval ends ($y=0$ and $%
y=\pi $) are associated with
\begin{eqnarray}
r_{p} &=&0\qquad p=1,\ldots ,n, \\
r_{0} &=&r.
\end{eqnarray}%
Identifying such real parameters with the mass of the involved modes, this
singular limit could generate massless ones. In this way, the $E_{8}$
symmetry factor appearing in many stringy models can be approached by taking
the following choice of weighted radius parameters
\begin{equation}
\delta _{0}=1,\quad \delta _{1}=2,\;\;\delta _{2}=4,\;\;\delta
_{3}=6,\;\;\delta _{4}=5,\;\;\delta _{5}=4,\;\;\delta _{6}=3,\;\;\delta
_{7}=2,\;\;\delta _{8}=3.
\end{equation}%
This can give rise to Higgsing field contents in M-theory
compactifications providing a possible geometrical evidence for the
existence of certain gauge theories relying on real toric fibrations.
Indeed, according to such a basic representation, the Dynkin weights
correspond to simple roots of the algebra. Removing a node from the
extended Dynkin diagram, it  can be viewed as the reducing symmetry pattern in
accord with adjoint Higgsing. Roughly, for the Dynkin diagrams
representation of the $E_{n}$ Lie algebras, the Higgsing i.e., the
corresponding symmetry breaking $E_{n}\rightarrow E_{n-1}$ can be
pictured as the suppression of a particular node from the $E_{n}$
Dynkin diagram  \cite{201,202,203}. The depiction that has to be employed
for the Higgsing is the one with the suppressed node, whose the highest
weight is equal to its corresponding root. This could be approached using
quiver gauge theories based on affine Dynkin diagrams.

An examination reveals that one could anticipate a possible conjectured
duality formulated as follows
\begin{equation*}
\mbox{ M-theory  on }\mathbf{R}^{1,8}\times \mathbb{T}^{2}\times \lbrack
0,\pi ]\longrightarrow \mbox{ Heterotic     theory  on }\mathbf{R}%
^{1,7}\times \mathbb{T}^{2}
\end{equation*}%
where one circle of heterotic string theory is identified with the
non-vanishing circle of M-theory real torus.   One could  provide certain arguments in favor of this anticipated mapping duality. At first sight, this duality seems a little  bit  different since generally one exploits complex structures to approach certain dualities  associated with F-theory compactifications. However, here, we are implementing   real parameters. The latters, which have been not explored, could unveil new data on the anticipated duality. In addition to this, the adiabatic argument could also bring extra supports \cite{adiaba}. Since M-theory on $\mathbb{T}^{2}$ is equivalent to type IIB on $S^1$, we can compactify both sides on the interval by varying the size parameters. This could be linked to type I superstring \cite{duality}. In certain limits, the latter is dual to heterotic string in 9-dimensions. Based on these arguments, one should expect that there could be  such a conjectured duality as the one appearing in F-theory relying on shape parameters. Forgetting about other arguments, this disinterest could be  due to the lack of mathematical  results associated with real fibrations. This needs more reflections
which are left for future works.

\section{Extended compactification models}
In this section, we would like to generalize the previous finding.
Concretely, we consider a generalized  real toric fibration. A close inspection
shows that the previous real toric fiber can be extended as follows
\begin{equation}
S^{1}\times S^{1}\longrightarrow \mathbb{\mathcal{T}}^{n}=S^{1}\times
S^{n-1}.
\end{equation}%
where $S^{1}$ will be replaced by $S^{n-1}$. It is worth noting that the
real toric geometry $S^{1}\times S^{n-1}$ has been approached in connection
with black ring solutions \cite{HD}. A special emphasis has been put on $n=3$
associated with black rings in five dimensions considered as an intermediate
solution between black holes and black strings \cite{BR}. Roughly speaking, $%
\mathcal{T}^{n}=S^{1}\times S^{n-1}$ can be described by the following
algebraic equation
\begin{equation}
\left( R-\sqrt{x_{n+1}^{2}+x_{n}^{2}}\right)
^{2}+\sum_{i=1}^{n-1}x_{i}^{2}=r^{2},  \label{41}
\end{equation}%
where one recovers (\ref{31}) by taking $n=2$. In this way, $\mathcal{T}^{n}$
reduces to $\mathbb{T}^{2}$. An examination reveals that the above
parametrization could be generalized in different ways depending on the
fibration of $S^{1}$ over the base geometry $S^{n-1}$. In the generic
fibration, the size of the fiber $S^{1}$ should be, in principle, a function
of the local coordinates of $S^{n-1}$ and its size parameters as follows
\begin{equation}
R=R(r,\theta _{1},\ldots ,\theta _{n-1}),
\end{equation}%
constrained by the toric real geometry (\ref{41}). For $n>2$, we could deal
with two relevant configurations. First, we consider a simple fibration,
where the fiber depends only on one local coordinate of $S^{n-1}$. In this
way, the fiber size can take the following form
\begin{equation}
R=R+f_{s}(r,\theta _{\ell }).
\end{equation}%
The second relevant configuration corresponds to a generic fibration
function depending on all local coordinates of $S^{n-1}$ and its size
parameters. In this case, we can write
\begin{equation}
R=R+f_{g}(r,\theta _{i}),\quad i=1,2,\ldots ,n-1.
\end{equation}%
In what follows, we will be interested in the second situation being a
generic one. In particular, we propose the following parametrization
\begin{eqnarray}
x_{1} &=&r\cos \theta _{1},  \notag \\
&&\vdots  \notag \\
x_{n-1} &=&r\cos \theta _{n-1}\sin \theta _{1}\ldots \sin \theta _{n-2}, \\
x_{n} &=&(R+r\sin \theta _{1}.\ldots \sin \theta _{n-1})\cos \theta _{n},
\notag \\
x_{n+1} &=&(R+r\sin \theta _{1}.\ldots \sin \theta _{n-1})\sin \theta _{n},
\notag
\end{eqnarray}%
where we have taken $f_{g}$ as
\begin{equation}
f_{g}(r,\theta_{i})=f_{g}=r\prod_{i=1}^{n-1}\sin \theta _{i}.
\end{equation}%
It is worth noting that for $n=2$ the equation (\ref{n}) can be recovered by
shifting $\theta _{1}$ as follows
\begin{equation}
\theta _{1}\longrightarrow \theta _{1}-\frac{\pi }{2}.
\end{equation}%
In M-theory compactifications down to eight dimensions, the geometry $%
\mathbb{T}^{2}\times I$ will be replaced by $\mathcal{T}^{n}\times I$ where
the size parameter $R$ is replaced, as before, by $R\sin \eta $. It is
remarked that at the end of the interval similar singular behaviors appear
associated with the existence of $n$-dimensional sphere $S^{n}$ with radius $%
r$. A connection with Lie symmetries requires certain constraints on such
compactifications. Indeed, it is recalled that the Euler caracteristic of $%
S^{n}$ is given by
\begin{equation}
\chi (S^{n})=1+(-1)^{n}.
\end{equation}%
Diagonal elements of the Cartan matrices required by
\begin{equation}
\chi (S^{n})=S^{n}\times S^{n}=2,
\end{equation}%
being the self intersection, impose that $n$ should be even
\begin{equation}
n=2m.
\end{equation}%
In the singular behaviors of the M-theory compactification, $S^{1}\times
S^{2m-1}$ can be viewed as a collection of $S^{2m}$ according to affine
Dynkin diagrams providing similar findings obtained in the previous section.

\section{Concluding remarks and open questions}

The moduli space of $\mathbb{T}^{2}$ involves two different kinds of
parameters which control the corresponding shape and size. The shape
deformations have been extensively used in F-theory complex analysis by
ignoring the other part of the moduli space associated with size
deformations. Motivated by such activities, we have investigated   M-theory compactifications
from the size region of such a moduli space. Concretely, we have presented a
road to approach M-theory compactifications relying on real parameters
controlling the size of internal spaces. Concretely, we have implemented a
real algebraic geometry of $\mathbb{T}^{2}$ in order to explore the physics
associated with the size contributions. In particular, we have provided
certain compactifications from real fibrations over real bases. Precisely,
we have discussed, in some detail, eight dimensions obtained from the
interval compacatification showing two different relevant behaviors. In
particular, we have revealed that the middle of such an interval is
associated with the torus geometrical behaviors, while the ends correspond
to singular ones, discussed in terms of the Euler characteristic changing
over the real base. Examining such behaviors, we have approached the gauge
Higgsing content by combining many results explored in string theory and
related models.

This work comes up with many remarks and open questions related to such real
fibration compactifications. Assuming its existence, one natural question is
to establish possible links with certain limits of known supergravity
theories using duality symmetries. We anticipate to stress that there could
be some links with theories based on real analysis in M-theory
compactfications including G2 manifolds. Moreover, it is worth noting that
such geometrical features and tools of real torus fibrations could be used
in the explicit building of models with a gauge group and spectrum
resembling that of the Standard Model after  possible compactification
scenarios \cite{30}. Principally, the most difficult question that the
stringy-inspired phenomenology encounters resides in the appropriate
compactification manifolds. In fact, it is just the topology of such a
manifold that yields important information about the the low-energy physics,
i.e. the number of supersymmetries and the fields content that survive the
compactification \cite{32}. In connection with the main theme in this work
concerned with real toric fibrations, one has to deal with a geometric
description of elliptically fibered spaces where a higher dimensional
manifolds with only free size deformation parameters.  A concrete view of the relevance of such a
geometry appears in intersecting D-brane model buildings. Especially,
realizations in orientifold compactifications of type II string theory with
D-branes where the extra dimensions are wrapped on certain homology cycles
in real toric fibrations, being the simple one where the homology cycles are
easily viewable giving a non-trivial physics. Actually, it is the rich
vocabulary such as intersection numbers $I$, intersection angles $\varphi
_{i}$, cycles $\pi _{i}$ and winding numbers $(n_{i},m_{i})$, between the
toric real geometry and the effective field theory existing in any D-brane
model that is behind the purposes for the prominence of D-brane physics \cite%
{34,35}. We thus believe that the corresponding realistic physics from real
torus fibrations still merits a careful and deep analysis. We hope to  address
such questions in future investigations.

\section*{Acknowledgements}

The authors would like to thank M. P. Garcia del Moral, A. Marrani and E.
Torrente Lujano for discussions on related topics and correspondence.  They thank also the
anonymous referee for its  careful reading of  the present  manuscript, insightful comments, and suggestions,  improving  this  work significantly. This
work is partially supported by the ICTP through AF.

\end{document}